\documentclass[%
reprint,
superscriptaddress,
nofootinbib,
amsmath,amssymb,
prl,
]{revtex4-1}

\usepackage{graphicx}
\usepackage{dcolumn}
\usepackage{bm}

\usepackage{graphicx}
\usepackage{dcolumn}
\usepackage{bm}
\usepackage{amsmath}
\usepackage{braket}
\usepackage{xr}
\usepackage{tabularx}
\usepackage{array}
\usepackage{siunitx}
\newcolumntype{.}{D{.}{.}{-1}}
\newcolumntype{q}{D{,}{.}{-1}}
\newcolumntype{d}[1]{D{.}{.}{#1} }

\usepackage{hyperref}

\begin{document}


\title{Revealing Color Forces with Transverse Polarized Electron Scattering}

\newcommand{\ANL}{Argonne National Laboratory, Argonne, IL}
\newcommand{\Temple}{Temple University, Philadelphia, PA}
\newcommand{\Tohoku}{Tohoku University, Tohoku, Japan}
\newcommand{\IHEP}{Kurchatov Institute - IHEP, Protvino, Moscow region, Russia}
\newcommand{\JLab}{Thomas Jefferson National Accelerator Facility, Newport News, VA}
\newcommand{\deceased}{Deceased.}
\newcommand{\MSU}{Mississippi State University, Starkville, MS}
\newcommand{\UVA}{University of Virginia, Charlottesville, VA}

\author{W.~Armstrong}\affiliation{\Temple}\affiliation{\ANL}
\author{H.~Kang}\affiliation{Seoul National University, Seoul, Korea}
\author{A.~Liyanage}\affiliation{Hampton University, Hampton, VA}
\author{J.~Maxwell}\affiliation{\UVA}\affiliation{\JLab}
\author{J.~Mulholland}\affiliation{\UVA}
\author{L.~Ndukum}\affiliation{\MSU}
\author{A.~Ahmidouch}\affiliation{North Carolina A\&M State University, Greensboro, NC}
\author{I.~Albayrak}\affiliation{Hampton University, Hampton, VA}
\author{A.~Asaturyan}\affiliation{Yerevan Physics Institute, Yerevan, Armenia}
\author{O.~Ates}\affiliation{Hampton University, Hampton, VA}
% ...

\collaboration{SANE Collaboration}

\date{\today}

\begin{abstract}
The Spin Asymmetries of the Nucleon Experiment (SANE) measured two double spin 
asymmetries using a polarized proton target and polarized electron beam at two 
beam energies, 4.7 GeV and 5.9 GeV. A large-acceptance open-configuration 
detector package identified scattered electrons at 40$^{\circ}$ and covered a 
wide range in Bjorken $x$ ($0.3 < x < 0.8$). Proportional to an average color 
Lorentz force, the twist-3 matrix element, $\tilde{d}_2^p$,  was extracted from 
the measured asymmetries at $Q^2$ values ranging from 2.0 to 6.0 GeV$^2$. The 
data display the opposite sign compared to most quark models, including the 
lattice QCD result, and an apparently unexpected scale dependence. Furthermore 
when combined with the neutron data in the same $Q^2$ range the results suggest 
a flavor independent average color Lorentz force.
\end{abstract}

\maketitle


Today, it is accepted that Quantum Chromodynamics (QCD), the gauge theory of 
strong interactions, plays a central role in our understanding of nucleon 
structure at the heart of most visible matter in the universe. QCD successfully 
describes many observables in high energy scattering processes where the 
coupling among the confined constituents of hadrons (quarks and gluons) is 
weak and perturbative (pQCD) calculations are possible, taking advantage of 
factorization theorems and evolution equations similar to quantum 
electrodynamics (QED). At the same time, QCD offers a clear path to unravel the 
non-perturbative structure of hadrons using lattice QCD, a powerful {\it ab 
initio} numerical method that provides the best insight when the coupling among 
the constituents is strong. 

The most fascinating property of QCD is confinement, which must arise from the 
dynamics of the partons inside hadrons. A small window into this dynamical 
behavior is offered by observables sensitive to quark-gluon correlations (providing confining forces) inside 
the spin-$\tfrac{1}{2}$ nucleon. An operator product expansion (OPE) provides 
well-defined quantities which codify not only the well known parton 
distributions in the nucleon, but also quark-gluon correlations lacking a naive 
partonic interpretation. Taking advantage of the spin-$\tfrac{1}{2}$ nucleon, 
these quantities can be measured in polarized inclusive deep inelastic electron 
scattering experiments and calculated as well, using lattice QCD (for review 
see~\cite{Jaffe:1996zw}).

The principal focus of this Letter is the measurement of the dynamical twist-3 
matrix element, $\tilde{d}_2$, which is interpreted as an average transverse 
color Lorentz force~\cite{Burkardt:2008ps} a quark feels due to the remnant at the space-time point
just as it is struck by the virtual photon.
Most importantly, a transversely polarized nucleon target probed with polarized 
electrons yields a unique experimental situation, where this color 
Lorentz force can be directly measured and used to test {\it ab initio} lattice QCD calculations. 

This semi-classical interpretation of $\tilde{d}_2$ as an average transverse color Lorentz force
is valid in the infinite momentum frame (IMF) of the proton, which is moving with velocity $\vec{v} = -c\hat{z}$.
Using light-cone variables, the $\hat{y}$-component of the Lorentz force acting on a color charge $g$
moving in the IMF is
\begin{equation}\label{eq:gluonFSTensor}
  g\,\left[ \vec{E} + \vec{v} \times \vec{B}\right]^{y} 
  = g\,\left[E_y + B_x\right] 
  = g\,G^{+y}
\end{equation}
where $G^{+y}$ is a component of the gluon field 
strength tensor.
Appearing in the definition of the local matrix element, $G^{+y}$ 
connects $\tilde{d}_2$ to the semi-classical transverse force interpretation 
\begin{equation}
  \begin{split}
    F^y(0) &\equiv  -\frac{\sqrt{2}}{2P^{+}} \langle P,S 
    \left|\bar{q}(0)G^{+y}(0)\gamma^{+}q(0)\right|P,S\rangle\\
    &= -\sqrt{2} M P^{+}S^x \tilde{d}_2 = -M^2 \tilde{d}_2~,
  \end{split}
\end{equation}
where the last equality is only valid in the proton's rest frame.

%
How do we access $\tilde{d}_2$? The nucleon spin structure functions, $g_1$ and $g_2$, parameterize the 
asymmetric part of the hadronic tensor in inclusive electromagnetic scattering, 
which through the optical theorem, is 
related to the forward virtual Compton scattering amplitude, $T_{\mu\nu}$. The 
reduced matrix elements of the quark operators appearing in the OPE analysis of 
$T_{\mu\nu}$ are related to Cornwall-Norton (CN) moments~\cite{Cornwall:1968cx} of the spin structure 
functions. At next-to-leading twist, the CN moments give 
\begin{align}\label{eq:g1moments}
   \displaystyle
   \int_0^1 x^{n-1} g_1(x,Q^2) dx &= a_n + 
   \mathcal{O}\big(\frac{M^2}{Q^2}\big), && n = 1,3,\dots
\end{align}
and
\begin{align}\label{eq:g2moments}
\begin{split}
   \displaystyle
   \int_0^1 x^{n-1} g_2(x,Q^2) dx &= \frac{n-1}{n}(d_n - a_n) + 
   \mathcal{O}\big(\frac{M^2}{Q^2}\big), \\
   &\quad\quad\quad n = 3,5,\dots
\end{split}
\end{align}
where $a_n = \tilde{a}_{n-1}/2$ and $d_n = \tilde{d}_{n-1}/2$ are the twist-2 
and twist-3 reduced matrix elements, respectively, which for increasing values 
of $n$ have increasing dimension and spin\footnote{The twist of an operator is 
  equal to its dimension minus its spin, and in QCD is a measure of the degree 
  of interactions between the constituents of hadrons, with higher twist index 
  representing increased correlations, e.g. the lowest twist, twist-2,
  corresponds to asymptotically free quarks;
  twist-3 is a quark-gluon-quark ($qgq$) correlation; twist-4 is some
  permutation of $qqgg$ correlations, etc. See~\cite{Jaffe:1996zw} for a 
review.}. $M$ is the nucleon mass and $Q^2 =-q^2$, where $q^{\mu}$ is the 
four-momentum transfer.

Neglecting target mass corrections (TMCs), \textit{i.e.} $M^2/Q^2\rightarrow0$, the twist-3 matrix element can 
be extracted from the $n=3$ CN moments
\begin{equation}\label{eq:d2tildedef}
   \begin{split}
      \tilde{d}_2 &= \int_{0}^1 x^{2} \big( 3 g_T(x) - g_1(x) \big) dx\,
   \end{split}
\end{equation}
where $g_T = g_1 + g_2$.
The equation above shows how experimental access to $\tilde{d}_2$ is achieved through measurements of the spin structure functions $g_1$ and $g_2$.

The famous Wandzura-Wilczek (WW) relation~\cite{Wandzura:1977qf}, 
$g_2^{WW}(x)=-g_1(x) +\int_x^1 g_1(y) dy/y$, allows us to write
\begin{equation}\label{eq:gTWW}
  g_T(x) = \int_x^1 \frac{dy}{y}g_1(y) + \bar{g}_2(x)\,, 
\end{equation}
such that $\bar{g}_2$ contains the higher twist contribution to the $g_2$ spin structure function.
In the limit of vanishing quark mass  
(\ref{eq:d2tildedef}) can be evaluated using (\ref{eq:g1moments}) and (\ref{eq:g2moments}).
In this limit, $\bar{g}_2$ contains only dynamical higher-twist contributions.
At finite quark mass the WW relation still holds~\cite{Blumlein:1998nv}, however, $g_T$ picks-up
terms from the (twist-2) transversity parton distribution. These transversity contributions are the 
subject of recent theoretical investigations \cite{Accardi:2009au,Accardi:2017pmi}.

Nachtmann moments should be used at low $Q^2$ instead of CN moments as 
is emphasized in \cite{Dong:2008zg}. Definitions of the Nachtmann moments, 
$M_1^n$ and $M_2^n$, are found 
in \cite{Matsuda:1979ad,Piccione:1997zh,Dong:2008zg} where they appear as 
more complicated versions of equations (\ref{eq:g1moments}) and (\ref{eq:g2moments})
which mix $g_1$ and $g_2$. They are related to the reduced matrix elements through
\begin{align}
   M_1^{(n)}(Q^2) =a_{n} &= \frac{\tilde{a}_{n-1}}{2}, && \text{for}~n=1,3...  
   \\
   M_2^{(n)}(Q^2) =d_{n} &= \frac{\tilde{d}_{n-1}}{2}, && \text{for}~n=3,5...
\end{align}
where we use the convention of Dong\footnote{Some authors define the matrix 
   elements excluding a factor of 
   1/2~\cite{Kodaira:1978sh,Kodaira:1979pa,Kodaira:1979ib,Matsuda:1979ad}, 
   and/or use even $n$ for the moments~\cite{Jaffe:1990qh,Blumlein:1998nv}. In 
   this work  we use the convention of~\cite{Piccione:1997zh,Dong:2008zg} which 
   absorbs the 1/2 factor into the matrix element and use odd $n$ for the 
   moments, whereas, the matrix elements excluding the 1/2 and even $n$ are 
   $\tilde{a}_{n-1}$ and $\tilde{d}_{n-1}$.}. 
Nachtmann moments, by their construction, project out matrix elements of definite 
twist and spin, therefore, they do not contain any  $\mathcal{O}\big(\frac{M^2}{Q^2}\big)$ terms.
When the target mass is neglected these equations 
reduce to $M_1^1=\int g_1 dx$ and $2M_2^3 = \int x^2(2g_1+3g_2)dx$.

Because both twist-2 and twist-3 operators contribute at the same order 
in transverse polarized scattering, a measurement of $g_2$ provides \emph{direct} 
access to higher twist effects\cite{Jaffe:1989xx} and thus the force we are seeking in this measurement.
This puts polarized DIS in an entirely unique situation to test lattice QCD 
\cite{Gockeler:2005vw}~and models of higher twist effects.

The Spin Asymmetries of the Nucleon Experiment (TJNAF E07-003) was conducted at 
Thomas Jefferson National Accelerator Facility in Hall-C during the winter of 
2008-2009 using a longitudinally polarized electron beam and a polarized proton 
target. Inclusive inelastic scattering
data in both the deep inelastic scattering and nucleon resonance regions were taken with two 
beam energies, $E=4.7$ and $5.9$~GeV, and with two target polarization 
directions: longitudinal, where the polarization direction was along the 
direction of the electron beam, and transverse, where the target polarization 
pointed in a direction perpendicular to the electron beam. 
The polarization angle with respect to the electron beam was 80$^{\circ}$ for the 
transverse configuration in order to match the acceptance and kinematics of scattered
electrons in the longitudinal target configuration.
Scattered electrons were detected 
in a new detector stack called the big electron telescope array (BETA) and also 
independently in Hall-C's high momentum spectrometer (HMS). Here, we give a 
brief discussion of the experimental apparatus and techniques, which are 
discussed in more details in an instrumentation paper \cite{Maxwell:2017mhs}.

The beam polarization was measured periodically using a M\o{}ller polarimeter 
and production runs had beam polarizations from 60\% up to 90\%. The beam 
helicity was flipped from parallel to anti-parallel at 30 Hz and the helicity 
state, determined at the accelerator's injector, was recorded for each event.

A polarized ammonia target acted as an effective polarized proton target and 
achieved an average polarization of 68\% by dynamic nuclear polarization in a 5 
T field. NMR measurements, calibrated against the calculable thermal 
equilibrium polarization, provided a continuous monitor of the target 
polarization. To mitigate local heating and depolarizing effects, the beam 
current was limited to 100 nA and a raster system moved the beam in a 
$1$~cm radius spiral pattern. 
By adjusting the microwave pumping 
frequency, the proton polarization direction was reversed. These two directions, 
positive and negative target polarizations, were used to estimate associated 
systematic uncertainties, since taking equal amounts of data with alternating positive 
and negative target polarization largely cancels any correlated behavior in 
the sum. 

BETA consisted of four detectors: a forward tracker placed close to the target, 
a threshold gas Cherenkov counter, a Lucite hodoscope, and a large 
electromagnetic calorimeter called BigCal. BETA was placed at a fixed central 
scattering angle of 40$^\circ$ and covered a solid angle of roughly 200 msr.  
Electrons were identified by the Cherenkov counter, which had an average signal 
of roughly 18 photoelectrons\cite{Armstrong:2015csa}. The energy was determined 
by the BigCal calorimeter, which consisted of 1744 lead glass blocks placed 3.35 
m from the target. BigCal was calibrated using a set of $\pi^0 \rightarrow 
\gamma\gamma$ events. The Lucite hodoscope provided additional timing and 
position event selection cuts and the forward tracker was not used in the 
analysis of production runs. 

The 5 T polarized-target magnetic field caused large deflections for charged 
particle tracks. In order to reconstruct tracks at the primary scattering 
vertex, corrections to the momentum vector reconstructed at BigCal were 
calculated from a set of neural networks that were trained with simulated data 
sets for each configuration.

The invariant mass of the unmeasured final state is $W^2=M^2 + 2M\nu - Q^2$,
where $M$ is the proton mass, $\nu=E-E^{\prime}$ is the virtual photon energy, 
and $Q^2 = - q^2= 2EE^{\prime}(1-\cos\theta)$. The scattered electron energy 
($E^{\prime}$)  and angle ($\theta$) are used to calculate the Bjorken 
variable $x=Q^2/2M\nu$.
BETA's large solid angle and open configuration allowed a broad kinematic range 
in $x$ and $Q^2$ to be covered in a single setting.

The measured double spin asymmetries for longitudinal ($\alpha=180^{\circ}$) 
and transverse ($\alpha=80^{\circ}$) target configurations were formed using 
the yields for beam helicities pointing along ($+$) and opposite ($-$) the 
direction of the electron beam,
\begin{equation}
   \label{eq:asym_fpbpt}
  A_{m}(\alpha) \equiv \frac{1}{f(W,Q^{2}) P_B P_T} \Bigg[ 
  \frac{N_{+}(\alpha)-N_{-}(\alpha)}{N_{+}(\alpha)+N_{-}(\alpha)}\Bigg]\,.
\end{equation}
The normalized yields are $N_{\pm} = n_{\pm}/(Q_{\pm}L_{\pm})$,
where $n_{\pm}$ is the raw number of counts for each run ($\sim$~1 hour of beam 
on target), $Q_{\pm}$ is the accumulated charge for the given beam helicity 
over the counting period, and $L_{\pm}$ is the live time for each helicity, 
$f(W,Q^{2})$ is the target dilution factor, and the beam and target 
polarizations are $P_B$ and $P_T$ respectively. 
The target dilution factor, taking into account scattering from 
unpolarized nucleons in the target and depending on the scattered electron
kinematics, is discussed in detail in~\cite{Maxwell:2017mhs}.

The dominant source of background for this experiment came from the decay of 
$\pi^0$s into two photons which, subsequently, produce electron-positron pairs 
which are then identified as DIS electrons. A pair produced outside of the 
target no longer experiences a strong magnetic field deflection, and therefore 
the pair  travels in nearly the same direction. These events produced twice the 
amount of \v{C}erenkov light and are effectively removed with a cut in maximum 
signal amplitude\cite{Armstrong:2015csa}. However, pairs produced inside the target are 
sufficiently and oppositely deflected, causing BETA to observe only one particle 
of the pair. These events cannot be removed through selection cuts and are 
treated through a background correction.

The background correction was determined by fitting existing inclusive $\pi^0$ 
production data and running a simulation to determine their contribution 
relative to the measured inclusive electron scattering yields. The correction only becomes 
significant at scattered energies below 1.2 GeV, where the positron-electron 
ratio begins to rise. The background correction consisted of a dilution 
($f_{\text{BG}}$) and contamination ($C_{\text{BG}}$) term defined as
\begin{align}
   \label{eq:bgCorr}
   A_{b}(\alpha) &= A_{m}(\alpha)/f_{\text{BG}} - C_{\text{BG}}.
\end{align}
The contamination term was small and only increases to $~1\%$ at the lowest $x$ 
bin. The background 
dilution also increases at low $x$ and 
becomes significant ($>~10\%$ of the measured asymmetry) only for $x < 0.35$.

After correcting for the pair symmetric background, the radiative corrections 
were applied following the standard formalism laid out by Mo and Tsai 
\cite{Mo:1968cg} and the polarization dependent treatment of Akushevich, et al.  
\cite{Akushevich:1994dn}.
The elastic radiative tail was calculated from models of the proton form factor 
\cite{Arrington:2007ux}.
The pair-symmetric background-corrected asymmetry was then corrected with elastic 
dilution and contamination terms 
\begin{equation}
   \label{eq:A_el}
   \begin{split}
     A_{el}(\alpha) &= A_{b}(\alpha)/f_{el} - C_{el}\,,
   \end{split}
\end{equation}
where $f_{el}$ is the ratio of inelastic scattering to the sum of elastic and 
inelastic scattering, and $C_{el}$ is the polarized elastic scattering cross 
section difference over the total inelastic cross section.
The elastic dilution term remained less than 10\% of the measured asymmetry in 
the range $x=0.3$ to $0.8$ for both target configurations. In the same range of 
$x$, the longitudinal configuration's elastic contamination remained less than 
10\% in absolute value, whereas, the transverse configuration's elastic 
contamination remained less than a few percent in absolute units.

The last correction required calculating the polarization dependent inelastic 
radiative tail of the Born-level polarization-dependent cross sections, which 
form the measured asymmetry. However, numerical studies 
\cite{Mo:1968cg,Akushevich:1997di}
with various models indicate the size of this radiative tail is small for most 
kinematics, reaching a few percent only at the lowest and highest $E^{\prime}$ 
bins. More importantly, the contribution of this radiative tail to the 
inelastic asymmetry remains within the systematic uncertainties associated with 
the model and numerical precision of our calculations.  Therefore, this 
correction was treated as a systematic uncertainty. This situation can only 
improve with future precision measurements of the polarization-dependent cross 
sections by scanning beam energies at a fixed angle~\cite{Mo:1968cg}.


\begin{figure}
  \includegraphics[width=0.5\textwidth, trim=0mm 0mm 0mm 1mm,clip]{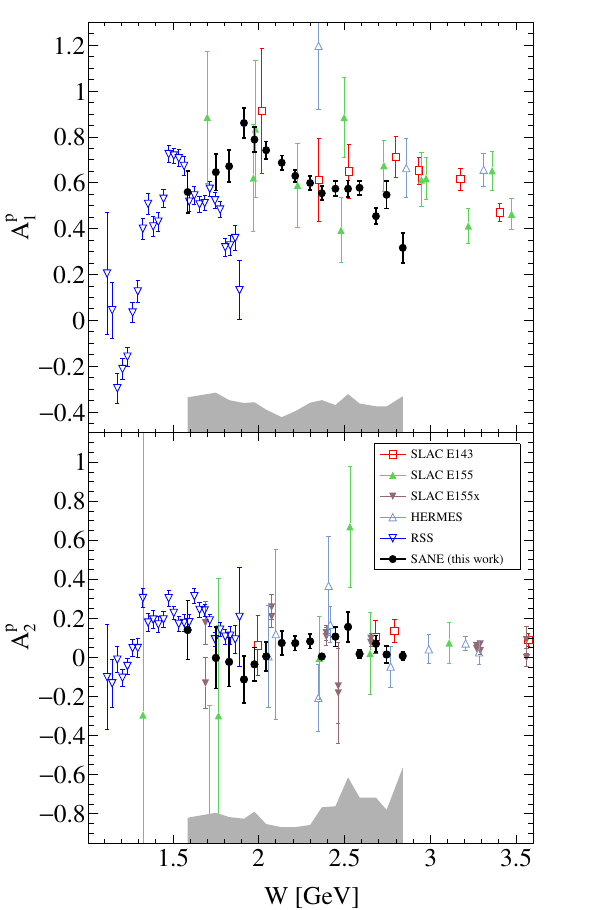}
  \caption{\label{fig:A1A2results}The SANE results (circle) and existing data 
    from SLAC's E143 (square)\cite{Abe:1996ag}, E155 (filled up triangle) 
    \cite{Anthony:1999py}  , E155x (filled down triangle)\cite{Anthony:2002hy}, 
    HERMES (up triangle) \cite{Airapetian:2011wu}, and $RSS$ (down triangle) 
    \cite{Slifer:2008xu} experiments for the virtual Compton scattering 
  asymmetries $A_1^p$ (top) and $A_2^p$ (bottom). The lower band shows systematic uncertainty. 
  Note the $A_1$ data shown are from experiments which measured both $A_{\parallel}$ and $A_{\perp}$. }
\end{figure}

The virtual Compton scattering asymmetries can be written in terms of the 
measured asymmetries
\begin{equation}
   \begin{split}
   \label{eq:A1extract}
     A_1 &= \frac{1}{D^{\prime} }
     \Big[\frac{E-E^{\prime}\cos\theta}{E+E^{\prime}}A_{180} \\
     &\quad\quad+ \frac{E^{\prime}\sin\theta}{(E+E^{\prime})\cos\phi} 
     \frac{A_{180}\cos\alpha + A_{\alpha}}{\sin\alpha} \Big]
   \end{split}
\end{equation}
and
\begin{equation}
   \begin{split}
   \label{eq:A2extract}
     A_2 &= \frac{\sqrt{Q^2}}{2 E D^{\prime}}\Big[ 
     A_{180}-\frac{E-E^{\prime}\cos\theta}{E^{\prime}\sin\theta\cos\phi}
     \frac{A_{180}\cos\alpha+A_{\alpha}}{\sin\alpha} \Big]
   \end{split}
\end{equation}
with $\alpha=80^{\circ}$ and where $A_{180}$ and $A_{80}$ are the corrected 
asymmetries, $D^{\prime}=(1-\epsilon)/(1+\epsilon R)$, 
$\epsilon=(1+2(1+\nu^2/Q^2)\tan^2(\theta/2))^{-1}$ is the degree of polarization of the longitudinal photon,
and $R=\sigma_L/\sigma_T$ is the ratio of longitudinal to 
transverse unpolarized cross sections. The combined results for $A_1$ and $A_2$ 
versus $W$ are shown in Fig.~\ref{fig:A1A2results}. These results significantly 
improve the world data on $A_2^p$.
The spin structure functions can be obtained from the measured asymmetries by 
using equations (\ref{eq:A1extract}) and (\ref{eq:A2extract}) along with
\begin{align}
   g_1 &= \frac{F_1}{1+\gamma^2}\big(A_1 + \gamma A_2\big)\\
   g_2 &= \frac{F_1}{1+\gamma^2}\big(A_2/\gamma - A_1 \big)\,,
\end{align}
where $\gamma^2=Q^2/\nu^2$ and $F_1$ is the unpolarized structure function.

Table~\ref{tab:moments} lists the measured moments and corresponding integrated $x$ 
ranges. The systematic uncertainties at the lower part of this range are dominated by 
the pair-symmetric background, which rapidly decreases towards higher $x$,
where the target polarization, 
target dilution, and beam polarization uncertainties are most significant.
Estimates for the low and high $x$ contributions and their uncertainties
were obtained from parton distribution fits to 
data~\cite{Bluemlein:2002be,Bourrely:2015kla,Sato:2016tuz} and fits to data in the 
resonance region~\cite{Drechsel:2007if}. In order to evaluate $\tilde d_2$ at a constant $Q^2$, 
evolution equations for $g_2$~\cite{Braun:2001qx} were used to estimate a correction 
for each $x$ point to provide $g_2$ at the same $Q^2$,
these corrections were found to be less than 1\% for nearly all $x$ points.
It is important to note that the moments include the point at $x=1$, which 
corresponds to elastic scattering on the nucleon. The elastic contributions to 
the moments are computed according to~\cite{Melnitchouk:2005zr} using empirical 
fits to the electric and magnetic form factors~\cite{Arrington:2007ux}. At large 
$Q^2$, the elastic contribution becomes negligible.

\begin{figure}[htb]
  \includegraphics[width=0.48\textwidth,clip,trim=3mm 2mm 22mm 14mm]{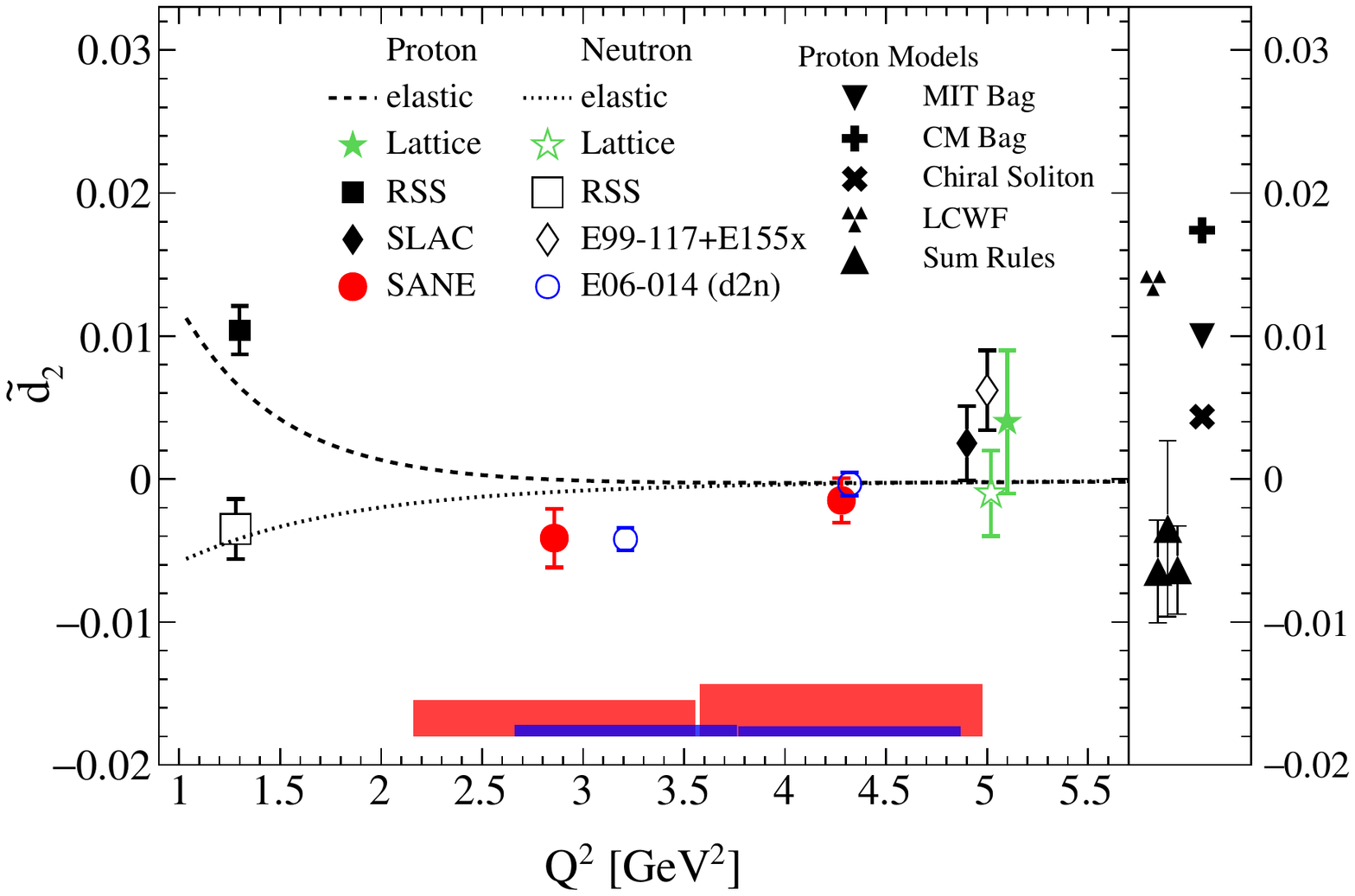}
  \caption{\label{fig:d2result}The results for $\tilde{d}_2$ of the proton from this work (SANE) and recent neutron results~\cite{Flay:2016wie} 
    with their systematic uncertainties (displayed in the lower bands). Also shown are the lattice QCD results
    \cite{Gockeler:2005vw}, previous proton (neutron) measurements with filled 
    (open) symbols from SLAC \cite{Anthony:2002hy}, E99-117 and 
    E155x~\cite{Zheng:2004ce}, and $RSS$~\cite{Wesselmann:2006mw,Slifer:2008xu} 
    experiments.
    The dashed (dotted) lines show the elastic contribution for the proton (neutron).
    The panel on the right shows proton model calculations from QCD sum rules \cite{Balitsky:1989jb,Stein:1994zk}, 
    the bag model~\cite{Signal:1996ct}, 
    the Center-of-Mass (CM) bag model \cite{Song:1996ea}, 
    the chiral soliton model \cite{Weigel:1996jh}, 
    and light-cone wave functions (LCWF) \cite{Braun:2011aw}. The models are calculated at $Q^2=5$~GeV$^2$, except the sum rules and LCWF, which were evaluated at $Q^2=1$~GeV$^2$.
}
\end{figure}

The results for the Nachtmann moment $2M_2^{(3)}(Q^2) = \tilde{d}_2(Q^2)$ are 
shown in Fig.~\ref{fig:d2result} along with a comparison to the two previous 
measurements, lattice QCD results, and model calculations. The first measurement 
was extracted from the combined results of the SLAC E143, E155, and E155x 
experiments~\cite{Anthony:2002hy}. The SLAC and lattice results are consistent 
with our result at $Q^2=4.3$~GeV$^2$.
The measurement from the Resonance Spin Structure ($RSS$) experiment (TJNAF 
E01-006)~\cite{
Slifer:2008xu}, extracted at $Q^2=1.28$~GeV$^2$, has a value 
$\tilde{d}_2^p = 0.0104\pm0.0016$, 
of which, $\sim 1/3$ comes from the inelastic contribution.

At both $Q^2=2.8$~GeV$^2$ and $Q^2=4.3$~GeV$^2$ our proton $\tilde d^p_2$ 
results are negative, although at $Q^2=4.3$~GeV$^2$ it is consistent with zero.  
Interestingly, when considered together with the world data, these results 
suggest a non-trivial scale dependence of $\tilde d_2$ --- positive at $Q^2\sim 
1$~GeV$^2$ as reported by $RSS$, becoming negative around $Q^2\sim 3$~GeV$^2$ 
as indicated by this work, finally, increasing slowly toward the positive SLAC 
measurement at $Q^2=5$~GeV$^2$ --- in contrast to the monotonic behavior 
expected from twist-3 pQCD evolution ~\cite{Shuryak:1981pi,Braun:2001qx}.
Furthermore, with the exception of the QCD sum rules, all model calculations and lattice QCD give positive 
values for the proton $\tilde{d}_2^p$. Intriguingly, our results 
complement a recent neutron $\tilde{d}_2^n$ 
measurement~\cite{Flay:2016wie,Posik:2014usi}, which shows a sizable negative 
value at $Q^2 \sim 3~$GeV$^2$, equal to that of the proton, as shown in 
Fig.~\ref{fig:d2result}. We note that while both experiments where performed at 
Jefferson Lab they used completely independent apparatus in two different 
Halls. Our proton results in combination with the world neutron results point 
to a flavor independent average color Lorentz force that has a puzzling apparent scale 
dependence in contrast with recent expectations~\cite{Burkardt:2008ps}.

\begin{table}
\centering
\caption{\label{tab:moments}Results for $\tilde{d}_2 = 2M_2^3 $ in units of $\times 10^{-3}$
with their statistical and systematic uncertainties. The low $x$, high $x$, and elastic systematic uncertainties were obtained from models. See text for details.}
\renewcommand{\arraystretch}{1.2}
\begin{ruledtabular}
  \begin{tabular}{ l
                 d{3.2} @{\hskip 0pt {$~$}\hskip -1pt} 
                 l      @{\hskip 1pt {}\hskip 0pt} 
                 d{1.2} @{\hskip 3pt {$\pm$}\hskip -5pt} 
                 d{1.2} @{\hskip 0pt {}\hskip 0pt}
                 d{4.2} @{\hskip 0pt {$~$}\hskip -1pt} 
                 l      @{\hskip 1pt {}\hskip 0pt} 
                 d{1.2} @{\hskip 3pt {$\pm$}\hskip -5pt} 
                 d{1.2} @{\hskip 0pt {}\hskip 0pt}
    }
    &  \multicolumn{4}{c}{$\langle Q^2 \rangle $ = 2.8 GeV$^2$ }  & \multicolumn{4}{c}{ $\langle Q^2 \rangle$ = 4.3 GeV$^2$ } \\
    $x_\text{low} - x_\text{high}$ & \multicolumn{4}{c}{$0.26 -  0.57$}   &  \multicolumn{4}{c}{$0.44 - 0.74$}   \\
    &  \multicolumn{4}{c}{\rule[4mm]{3cm}{0.5pt} }  & \multicolumn{4}{c}{ \rule[4mm]{3cm}{0.5pt} }\\[-5mm]
    &  \multicolumn{1}{c}{$~$}  & \multicolumn{2}{r}{\scriptsize (stat.)} &    \multicolumn{1}{c}{\scriptsize(sys.)} &       &      \multicolumn{2}{r}{\scriptsize(stat.)} &    \multicolumn{1}{c}{\scriptsize(sys.)} \\
    measured &    -4.77 & $\pm$  & 2.05 &    1.81 & -3.22 & $\pm$ & 1.56 & 3.57 \\
    low $x$  &     1.86 & $   $  &      &    0.13 &  2.47 & $   $ &      & 0.54 \\
    high $x$ &    -1.19 & $   $  &      &    1.81 & -0.49 & $   $ &      & 0.72 \\
    elastic  &    -0.04 & $   $  &      &    0.01 & -0.25 & $   $ &      & 0.02 \\
    total    &    -4.14 & $\pm$  & 2.05 &    2.56 & -1.49 & $\pm$ & 1.56 & 3.68 \\
\end{tabular}
\end{ruledtabular}
\end{table}


In summary, the proton's spin structure functions $g_1$ and $g_2$  have been 
measured at kinematics allowing for an extraction of $\tilde{d}_2$ at two different values 
of $Q^2$.  The present results in combination with the world data suggest an 
unexpected scale dependence of the average color Lorentz force and a flavor 
independence. Furthermore, precision measurements at 12 GeV Jefferson Lab with transversely polarized 
proton and neutron targets are justified to confirm this puzzling behavior~\cite{d2n12GeVHallC,solidSIDIS,solidSIDIS2}.
Moreover modern lattice QCD calculations of $\tilde{d}_2$, without the quenched 
approximation, which include disconnected diagrams~\cite{Gockeler:2005vw} , and 
are performed at the physical pion mass without chiral extrapolation, are sorely 
needed for a complete understanding of our observation. 

We would like to thank Vladimir Braun and Matthias Burkardt for discussions and 
feedback during the revision process. We would like to express our gratitude to the staff and technicians 
of Jefferson Lab for their support during the running of SANE.  We especially 
thank the Hall~C and Target Group personnel, who saw a technically challenging 
experiment through significant hardship to a successful end. This work was 
supported in part by the U.S.  Department of Energy, Office of Science, Office 
of Nuclear Physics, under contract no.  DE-AC02-06CH11357, DE-FG02-94ER4084, 
DE-FG02-96ER40950, and DE-AC05-06OR23177.

\bibliography{auto_refs}

\end{document}